\documentclass[prl,twocolumn,showpacs]{revtex4}
\usepackage{natbib}
\usepackage{epsfig}
\usepackage{amsmath,amssymb}
\usepackage{graphicx}
\usepackage{dcolumn}
\usepackage{bm}

\newcommand{\unit}[1]{\ensuremath{\,\mathrm{#1}}}
\newcommand{\Cd}{\ensuremath{^{111}\mathrm{Cd}^+}}
\newcommand{\state}[4]{\ensuremath{#1\,\,^{#2}\!{#3}_{#4}}}
\newcommand{\nm}{\unit{nm}}
\newcommand{\MHz}{\unit{MHz}}
\begin{document}

\title{Quantum Interference of Photon Pairs from Two Trapped Atomic Ions}
\author{P. Maunz}
\author{D. L. Moehring}
\author{M. J. Madsen}
\author{R. N. Kohn, Jr.}
\author{K. C. Younge}
\author{C. Monroe}
\affiliation{FOCUS Center and Department of Physics, University of Michigan, Ann Arbor, MI  48109}
\begin{abstract}
  We collect the fluorescence from two trapped atomic ions, and
  measure quantum interference between photons emitted from the ions.
  The interference of two photons is a crucial component of schemes to
  entangle atomic qubits based on a photonic coupling.  The ability to
  preserve the generated entanglement and to repeat the experiment
  with the same ions is necessary to implement entangling quantum
  gates between atomic qubits, and allows the implementation of
  protocols to efficiently scale to larger numbers of atomic qubits.
\end{abstract}
\maketitle

{ Trapped atomic ions are among the most attractive implementations of
  quantum bits (qubits) for applications in quantum information
  processing, owing to their long trapping lifetimes and long
  coherence times.  While nearby trapped ions can be entangled through
  their Coulomb-coupled motion \cite{Cirac95, Molmer99, Milburn00,
    Sackett00, Leibfried05, Haffner05}, it is more natural to entangle
  remotely-located ions through a photonic coupling, eliminating the
  need to control the ion motion.  When two atomic ions each emit a
  single photon \cite{blinov04,Volz06}, subsequent interference and
  detection of these photons can leave the trapped ion qubits in an
  entangled state \cite{Simon03, Duan04b}. Moreover, such a photonic
  coupling can be tailored to operate quantum gates between the ions
  and efficiently generate extended networks of entangled qubits and
  cluster states for scalable quantum computation \cite{Briegel98,
    Raussendorf01, Barrett05, Duan05, Lim05, Duan06}. Here, we report
  the generation of single photons and the observation of quantum
  interference between two photons emitted from two trapped atomic
  ions. The same two ions remain in place for the generation of
  thousands of two-photon interference events, which, together with
  the long coherence times available in trapped ion systems
  \cite{Bollinger91,Haffner05b,Langer05}, points the way toward
  scaling to large entangled networks of remotely-located qubits.}

Remote entanglement of two ions or atoms can be achieved by subjecting
two photons emitted by the particles to a Bell-state measurement and
is heralded by an appropriate coincidence detection of the
photons. The essence of this Bell-state measurement is the quantum
interference of two photons, which has been observed previously with
photons generated in a variety of physical processes and systems,
including nonlinear optical down-conversion
\cite{Hong87,Kaltenbaek06}, quantum dots \cite{Santori02}, atoms in
cavity-QED \cite{Legero04} and, more recently, two independently
trapped neutral atoms \cite{Beugnon06}.  We report the first
observation of interference between two photons emitted from multiple
trapped atomic ions.  This demonstration is important for scaling to
extended quantum networks of qubits, which is only feasible if the
entanglement can be preserved on a timescale long compared to the
average time needed to entangle two qubits.  With their unsurpassed
trapping and qubit coherence times, trapped ions are well suited for
this purpose \cite{Bollinger91,Langer05,Haffner05b}.

In the experiment, one or two \Cd\ ions are trapped in a four-rod
linear rf quadrupole trap with rod spacings of $0.5\unit{mm}$ and an
end-cap spacing of $2.6\unit{mm}$ \cite{Moehring06}. The rf drive
frequency is $\Omega_T/2\pi = 36\MHz$ and the center of mass secular
trapping frequencies are $(\omega_x, \omega_y, \omega_z)/2\pi =
(0.9,0.9,0.2) \MHz$. Residual micromotion at the rf drive frequency is
reduced by applying static offset voltages to the trap rods and
endcaps. Cadmium atoms from the background vapor are photoionized
using a frequency quadrupled ultrafast Ti:sapphire laser centered at
$229\nm$. The mean lifetime of the ions in the trap is over one hour.
%
%
Continuous wave (cw) laser light with a wavelength of $\lambda=214.5
\nm$ is used to Doppler cool and excite the ions. This light is
generated by frequency quadrupling the light from a cw amplified diode
laser at $858 \unit{nm}$ and is stabilized to a tellurium reference at
a detuning of $\Delta \approx -\Gamma/2 = -30 \MHz$ from the atomic
$\state{5s}{2}{S}{1/2} \leftrightarrow \state{5p}{2}{P}{3/2}$
transition of \Cd. Doppler cooling localizes the ion to about
$300\nm$, well outside the Lamb-Dicke limit but better than the
resolution of the diffraction-limited imaging optics. Incident
$\sigma^+$-polarized laser light optically pumps the ion to the $F=1,
m_F=1$ ground state and drives the closed optical transition to the
$F=2, m_F=2$ excited state, with the quantization axis defined by a
magnetic field of $0.5 \unit{Gauss}$.

Alternatively, the ions can be excited with ultrafast $\sigma^+$
polarized pulses, generated by a picosecond mode-locked Ti:sapphire
laser with a center frequency of $858 \nm$. An electro-optic pulse
picker is used to reduce the pulse repetition rate from $81 \MHz$ to
$27\MHz$ with an extinction ratio of better than 100:1 in the
infrared. Each of the pulses is then frequency quadrupled to
$214.5\nm$ through phase-matched LBO and BBO nonlinear crystals. The
UV (fourth harmonic) is filtered from the fundamental and second
harmonic via dichroic mirrors and directed to the ion with a near
transform-limited pulse duration of $1\unit{ps}$, exciting the ion on
a timescale much faster than the excited state lifetime of
$2.6\unit{ns}$ \cite{Moehring06}. The bandwidth of the pulsed laser
($\approx 0.4\unit{THz}$) is small compared to the fine-structure
splitting $(70\unit{THz})$, ensuring selective excitation to the
$\state{5p}{2}{P}{3/2}$ excited state.

\begin{figure}[htb]
\begin{center}
  \epsfig{file=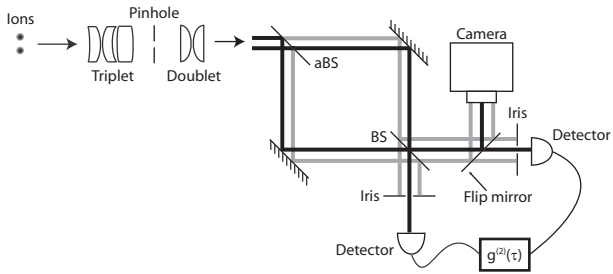,width=.93\columnwidth}
  \caption{Schematic of the detection system. The light from two ions is
    separated by an ancillary beam splitter (aBS) and superimposed on
    the primary beam splitter (BS). Both beam splitters are used at a
    $10^\circ$ opening angle. The non-overlapping ion images are
    blocked by two irises.  For a single ion and opened irises this
    system is equivalent to the Hanbury Brown and Twiss set-up. A flip
    mirror (FM) is used to send the beams on a single photon sensitive
    camera.}
\label{figSetup}
\end{center}
\end{figure}
The light scattered by the ions is collected using an objective lens
with a numerical aperture of $0.23$ and a working distance of $13
\unit{mm}$ (Fig.~\ref{figSetup}). The intermediate image generated
by the objective is re-imaged using a doublet lens, resulting in an
overall magnification of about 1000.
%
%
The light from both ions is sent on a $50\unit{\%}$ ancillary beam
splitter, then the transmitted and reflected beam pairs are directed
to the primary beam splitter where the light from the two different
ions is superimposed.
Behind this beam splitter, a removable mirror can be used to
send the light onto a camera which is used to monitor the ion
fluorescence during loading and for coarse alignment of the beams.  On
the camera, the ion images have a spot size of about $0.5 \unit{mm}$
and are separated by about $2 \unit{mm}$. Two irises are used to
select only the superimposed images of the ions. This light is
subsequently detected by two photon counting photomultiplier tubes
with a quantum efficiency of about $20\unit{\%}$ and a time resolution
of $1\unit{ns}$ \cite{Moehring06}.  The overall detection efficiency
of a photon emitted by an ion is $0.1\unit{\%}$. In the case of a
single trapped ion, the irises are opened and the set-up is equivalent
to a single beam splitter with two photodetectors.

This set-up is not sensitive to relative movement of the imaging
optics with respect to both ions and leads to a stable spatial overlap
of the modes even if the images of both ions move with respect to the
beam splitter. The equal path length of the beams of both ions
furthermore ensures that the modes match in size and wavefront radius
of curvature. From the interference contrast, the two pathlengths were
adjusted to match within $1\unit{mm}$.

\begin{figure}[htb]
\begin{center}
  \epsfig{file=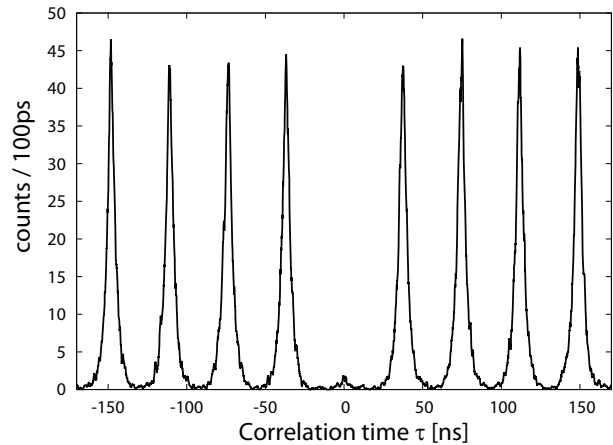,width=.93\columnwidth}
  \caption{Intensity autocorrelation of the light emitted by a single
    ion excited by one-picosecond pulses. The near-perfect
    antibunching at zero delay shows that at most one photon is
    emitted from an excitation pulse. The measurement was done with
    one ion with an excitation probability of about $20\unit{\%}$ from
    each excitation pulse. The data shown were integrated for about
    $5\unit{min}$ ($20\unit{min}$ total time).}
\label{figPulsedAutocorrelation}
\end{center}
\end{figure}

%
%
To demonstrate that the excitation of an ion with an ultrashort pulse
leads to the emission of at most one photon \cite{Darquie05}, we first
trap a single \Cd\ ion. We employ a repetitive sequence consisting of
a $150\unit{\mu s}$ cooling interval and a $50\unit{\mu s}$
measurement interval. During the cooling interval the ion is Doppler
cooled with cw-light only, and during the measurement interval the ion
sees only ultrafast laser pulses with a $37.5\unit{ns}$ pulse
separation. The intensity autocorrelation function of the photons
emitted during the measurement interval is recorded using a
multi-channel scaler and the resulting data are shown in
Fig.~\ref{figPulsedAutocorrelation}.
The periodic ultrafast excitation of the ion leads to peaks at
multiples of the pulse separation time of $37.5\unit{ns}$.  The half
width of these peaks is given by the $2.6\unit{ns}$ lifetime of the
excited state. In contrast to pulsed coherent or pulsed thermal light,
the peak at zero time delay is almost entirely suppressed. This
near-perfect antibunching is highly non-classical and demonstrates
that at most one photon is emitted after each excitation pulse. The
residual peak at zero time delay has a height of about $2\unit{\%}$ of
the other peaks and originates from diffusely scattered light of the
pulsed laser.  Theoretically, the probability to scatter two photons
from one ion excited with one pulse is limited by the emission
probability of an excited atom during the excitation pulse ($<
10^{-3}$ for our parameters).

Two-photon interference is a purely quantum phenomenon and can be
understood qualitatively if one considers the ways in which two
photons impinging on different input ports of a beam splitter can
emerge from separate output ports. There are two possibilities: both
photons are reflected, or both are transmitted. For two photons which
have the same polarization and frequency and which are exactly matched
on the beam splitter, these two cases interfere destructively, leading
to the effect that the photons always emerge together from the beam
splitter \cite{Mandel99,Legero05}. Here we demonstrate two-photon
interference by exciting two \Cd\ ions located in the same trap with
near-resonant cw light.

\begin{figure}[htb]
\begin{center}
\vspace{3ex}
  \epsfig{file=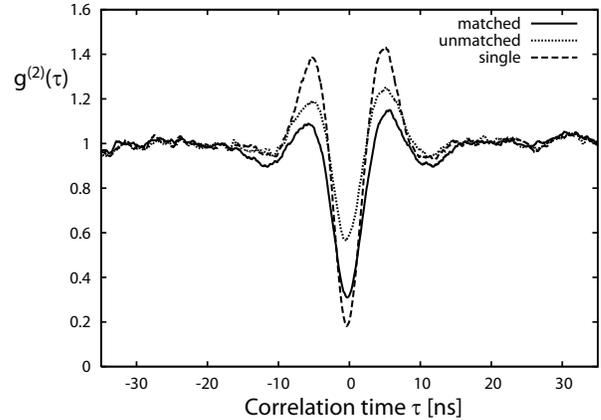,width=.90\columnwidth}
  \caption{Intensity autocorrelation and joint detection probability
    for cw-excitation (raw data), integrated for about $60\unit{minutes}$
    for each curve. Each ion leads to a total count rate of $4 \times
    10^{4}/\unit{s}$. The one ion intensity autocorrelation (dashed
    line) shows strong antibunching with $g^{(2)}( tau=0)=0.18$.  From
    this value we expect the joint detection probability at zero delay
    for light from two ions without mode overlap (dotted line) to be
    $0.59$, which is in good agreement with the experimental value of
    $0.57$. If the modes of the two ions are matched (solid line)
    two-photon interference leads to a significant reduction of
    coincidence detections. In this case the joint detection
    probability drops to $0.31$ at $\tau=0$.}
\label{figCWAutocorrelation}
\end{center}
\end{figure}

To separate the interference effect of photons emitted by different
ions from the emission properties of the individual ions using cw
excitation, we first investigate the photon statistics of a single ion
(dashed curve in Fig.~\ref{figCWAutocorrelation}). In this case, the
intensity autocorrelation function, $g^{(2)}(\tau)$, shows the well
known signal of damped Rabi oscillations \cite{Diedrich87,Itano88}:
when a photon is detected at time $t$, the atom is projected onto its
ground state and begins a damped Rabi oscillation which determines the
probability to observe a second photon at time $t+\tau$.
Theoretically, the probability of detecting two photons at the same
time vanishes, and background counts are measured to contribute less
than $1\unit{\%}$ of the signal. However, the instrument response
function of the photomultiplier tubes has a width of about
$1\unit{ns}$, limiting the antibunching observed in the
measurement. This antibunching means that the photons of one ion reach
the beam splitter one by one, however, the next photon of the same ion
can be detected after a short time.

We now investigate the joint detection probability of two photons for
light emitted by two ions, $P^{(2)}$. In this case, an observed joint
detection event can originate from two photons emitted successively by
one ion, $P^{(2)}_1=g^{(2)}$, or two photons emitted by different
ions, $P^{(2)}_2$. For a symmetric set-up --- equal emission rates of
both ions and a $50\unit{\%}$ beam splitter --- the joint detection
probability is
\begin{gather}
\label{theequation}
P^{(2)} = \frac{ g^{(2)} + P^{(2)}_2 }{2}.
\end{gather}

If the modes of the ions do not overlap on the beam splitter, the
photons from different ions are completely uncorrelated. However, due
to the single ion contribution to the joint detection probability, we
still observe anti-bunching, although of reduced depth, as can be
seen in Fig.~\ref{figCWAutocorrelation} (dotted line).
If the spatial modes of the two ions are matched on the beam
splitter, the photons reaching the beam splitter at the same time from
different ions interfere, thus reducing the number of coincidence
detections. This effect is clearly visible in the joint detection
probability depicted in Fig.~\ref{figCWAutocorrelation} (solid line).

\begin{figure}[htb]
\begin{center}
\vspace{3ex}
  \epsfig{file=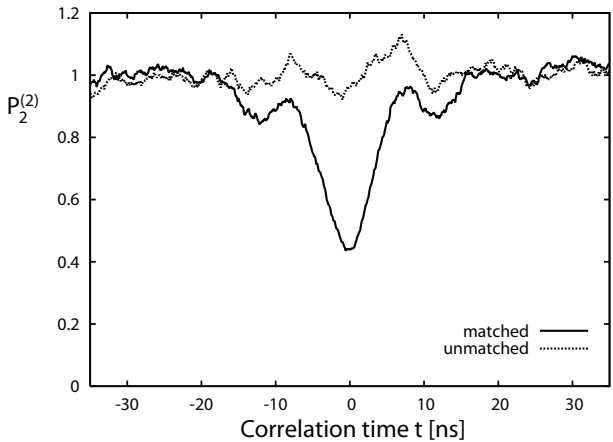,width=.93\columnwidth}
  \caption{Joint detection probability of photons emitted by different
    ions, $P^{(2)}_2$, as calculated from the measured single ion
    intensity autocorrelation, $g^{(2)}$, and the joint detection
    probability shown in Fig.~\ref{figCWAutocorrelation}. Without mode
    overlap (dotted line), the photons are uncorrelated and no
    anti-bunching is observed. When mode overlap is achieved (solid
    line), two-photon interference clearly reduces coincidence
    detections. The anti-bunching is expected to have a Gaussian shape
    where the depth is given by the mode overlap while the width is
    determined by the photon duration \cite{Legero05}. Our results
    correspond to a mode overlap of at least $57\unit{\%}$, and a
    photon duration of about $5.3\unit{ns}$.}
\label{figCWabAutocorrelation}
\end{center}
\end{figure}

Assuming a symmetric set-up, we can separate the two-photon
interference effect eminent in the joint detection probability of two
photons from different ions from the contribution of the single ion,
$g^{(2)}$, by solving for $P^{(2)}_2$ in eq. (\ref{theequation}). The
results are shown in Fig.~\ref{figCWabAutocorrelation}. 
For non-overlapping photon modes the joint detection probability
$P^{(2)}_2$ is basically flat and demonstrates that photons of
different ions are not correlated. If the modes of the two photons
overlap, we expect to find a Gaussian dip where the half width is given
by the duration of a photon and the depth is determined by the mode
overlap \cite{Legero05}. We measure a half width of about $5.3
\unit{ns}$ and a contrast of about $57\unit{\%}$, corresponding to a
mode-overlap of $57\unit{\%}$ ($75\unit{\%}$ amplitude matching).  We
attribute the non-perfect mode overlap in the experiment to phase
front distortions of the beams from the two ions.  These originate
mainly in the limited surface quality of the vacuum windows and the
lenses, considering the short radiation wavelength of $214 \unit{nm}$.

It is a challenge to realize satisfactory and stable mode overlap in
free space. While the set-up is suited to reject common mode movement of
the trap with respect to the beam splitter, changes in the ion
separation strongly affect the mode overlap. The resulting coincidence
detections lead to false positive events in the Bell-measurement and
thus limit the fidelity of any entanglement scheme.
Upon excitation of two ions with the pulsed laser we were not able to
demonstrate two-photon interference, even though an estimate shows
that heating of the ion by the pulsed laser should not be an
issue. Possible explanations are slight shifts in the separation of
the ions or degradation of the vacuum and heating of the ions due to
electrons released from the trap by the high-energy photons of the
pulsed laser. While our present mode overlap is sufficient to entangle
two ions, the rejection of other spatial modes, which can be achieved
with an single-mode optical fiber, should strongly improve the mode
overlap, producing a higher fidelity of the heralded entanglement
process.

In conclusion, we demonstrated a single photon source based on the
ultrafast excitation of a single trapped ion and we measured two
photon quantum interference of photons emitted by two trapped
ions. The contrast of the observed interference is sufficient to
demonstrate entanglement of two ions in this set-up. In the future,
single-mode fibers should greatly improve the mode
overlap and thus lead to the entanglement of remote ions with high
fidelity.  Building upon this, entangling gates provide a means to
scale the probabilistic entanglement from two qubits to the
generation of networks of entangled qubits.  The tremendous
advantages of the trapped ion system --- extremely long storage and
coherence times and high readout fidelity --- should make the scalable
entanglement of many qubits feasible.

\begin{acknowledgments}
  This work is supported by the National Security Agency and the Disruptive
  Technology Office under Army Research Office contract W911NF-04-1-0234, and
  the National Science Foundation Information Technology Research Program.
\end{acknowledgments}


\end{document}